%% file: hc_to_hadrons.tex
\RequirePackage{lineno}
\documentclass[aps,prd,twocolumn,showpacs,byrevtex]{revtex4-1}
\usepackage{times, amsmath, graphicx, color, overpic, enumitem}
\usepackage{multirow, lineno, verbatim, setspace, array}
\usepackage{ulem}
\usepackage[pdfstartview=FitH, colorlinks, urlcolor=blue, citecolor=blue,linkcolor=blue]{hyperref}
\usepackage{epstopdf}

\newcommand{\jpsi}{J/\psi} \newcommand{\psip}{\psi(3686)}
\newcommand{\br}{\mathcal{B}}\newcommand{\pizh}{\pi^{0}_{h}}\newcommand{\pizl}{\pi^{0}_{l}}
\newcommand{\pip}{\pi^{+}} \newcommand{\pim}{\pi^{-}} \newcommand{\piz}{\pi^{0}}
\newcommand{\pbar}{\bar{p}} \newcommand{\pb}{\bar{p}}
\newcommand{\kp}{K^{+}} \newcommand{\km}{K^{-}} 
 \newcommand{\etac}{\eta_{c}}
 
\newcommand{\csq}{c^{2}}\newcommand{\p}{p}\newcommand{\chisq}{\chi^{2}}
 \newcommand{\hc}{h_{c}}

\begin{document}

%\linenumbers
\DeclareGraphicsExtensions{.eps,.png,.ps}

\title{\boldmath{First observations of $\hc\to$hadrons}}

\input{authors.tex}

\begin{abstract}
  Based on $(4.48 \pm 0.03)\times10^{8}$ $\psip$ events, collected with the BESIII detector {\color{black}{at the BEPCII storage ring}}, five $\hc$ hadronic decays are searched for via the process $\psip \to \piz \hc$. Three of them, $\hc \to p \pbar \pip \pim$, $\pip \pim \piz$, and $2(\pip \pim) \piz$, are observed for the first time with significances of {\color{black}{7.4$\sigma$, $4.6\sigma$, and 9.1$\sigma$}}, and their branching fractions are determined to be $(2.89\pm0.32\pm0.55)\times10^{-3}$, $(1.60\pm0.40\pm0.32)\times10^{-3}$, and $(7.44\pm0.94\pm1.52)\times10^{-3}$, respectively, where the first uncertainties are statistical and the second systematic. No significant signal is observed for the other two decay modes, and the corresponding upper limits of the branching fractions are determined to be $\br(\hc\to3(\pip\pim)\piz)<8.7\times10^{-3}$ and $\br(\hc\to\kp\km\pip\pim)<5.8\times10^{-4}$ at the 90\% confidence level.
\end{abstract}
\maketitle

\par The study of charmonium states is crucial for reaching a deeper understanding of the low-energy regime of quantum chromodynamics (QCD), a theory describing the strong interaction, which has been tested successfully at high energy. Since its discovery in 2005~\cite{Rubin:2005px,Rosner:2005ry}, there have been few measurements of the decays of the spin-singlet charmonium state $h_c(^1P_1)$. Its best-measured decay is the radiative transition $h_c \to \gamma \eta_c$~\cite{e835, cleo-confirm, bes3-confirm}, while the sum of the other known $\hc$ decay branching fractions is less than $3\%$~\cite{pdg-2016}. Among these measurements, there is only evidence for one $h_c$ hadronic decay, $h_c \to 2(\pi^+\pi^-)\pi^0$, which was reported by CLEO-c with a statistical significance of $4.4\sigma$~\cite{Adams:2009aa}.

Improved measurements and observation of new $h_c$ hadronic-decay modes will shed light on the $\hc$ decay mechanism, and be helpful for guiding the development of QCD based models. For example, perturbative QCD (pQCD){\color{black}~\cite{pqcd-01, pqcd-02, pqcd-03}} and non-relativistic QCD (NRQCD){\color{black}~\cite{nrqcd-01, nrqcd-02, nrqcd-03}} are two alternative models for describing features of  low-energy QCD, and their predicted ratios of the hadronic width of the $\hc$ to that of the $\etac$ ($\Gamma_{\hc}^{\rm had} / \Gamma_{\etac}^{\rm had}$) are very different~\cite{theory-1}, as is the corresponding ratio involving decays of $\jpsi$ mesons ($\Gamma_{\hc}^{\rm had} / \Gamma_{\jpsi}^{\rm had}$).
New studies of $\hc$ hadronic decays will enable these ratios to be measured, and comparisons to be made with the theoretical predictions.

The discovery of $\hc$ hadronic decays provides new tag channels that can be used in XYZ  (charmonium-like) studies with $\hc$ as the intermediate state.  This would provide a boost in signal yield comparable to that available from the tag channel $\hc \to \gamma \eta_{c}$, $\etac \to$hadrons, which is the only mode applied at present.

Improved studies of $\hc$ decays can be made with the large $\psip$ sample of $4.48\times10^{8}$ events~\cite{psip-numerr}, {\color{black}{produced via $e^{+}e^{-}$ collisions}}, which has been collected with the BESIII detector. In this Letter, we report the first observations of decays $\hc \to p \pbar \pip \pim$, $\pip \pim \piz$, and $2(\pip \pim)\piz$, and upper limits of the branching ratios for the decays $\hc \to 3(\pip \pim)\piz$ and $\kp \km \pip \pim$.

%%%%%%%%%%%%%%%%%%%%%%%%%%%%%%%%%%%%%%%%%
% Data samples and MC                   %
%%%%%%%%%%%%%%%%%%%%%%%%%%%%%%%%%%%%%%%%%
\par The BESIII detector~\cite{detail_bepcii} is a general purpose detector with a 93\% solid angle coverage. A small-cell helium-based multi-layer drift chamber (MDC) determines the momentum of charged particles in a 1\,T magnetic field with a resolution of 0.5\% at 1\,GeV/$c$, and measures their ionization energy loss ($dE/dx$) with resolutions better than 6\%. A CsI(T1) electromagnetic calorimeter (EMC) measures the photon energies with resolutions 2.5\% (5.0\%) in the barrel (end caps). A time-of-flight system (TOF), composed of plastic scintillators with resolution of 80\, ps (110\,ps) in the barrel (end caps), is used for particle identification (PID). A resistive plate chambers based muon counter with 2\,cm position resolution is used for muon identification.

\par To obtain the detection efficiencies, signal Monte Carlo (MC) samples for the processes $\psip \to \piz \hc$, and $\hc \to \p \pb \pip \pim$, $\pip\pim\piz$, $2(\pip\pim)\piz$, $3(\pip\pim)\piz$, or $\kp\km\pip\pim$ are generated based on phase-space distributions. To investigate the background, an inclusive MC sample of $5.06\times10^{8}$ $\psip$ events is generated, in which the $\psip$ resonance is produced with {\sc kkmc}~\cite{kkmc_01, kkmc_02}. Decays with known branching fractions obtained from the Particle Data Group (PDG)~\cite{pdg-2016} are generated with {\sc evtgen}~\cite{evtgen}, while the other decays are generated with {\sc lundcharm}~\cite{lundcharm}. In all the simulations, the {\sc geant4}-based~\cite{geant4_01, geant4_02} package {\sc boost}~\cite{boost} is used to model the detector responses and to incorporate time-dependent beam backgrounds.

%%%%%%%%%%%%%%%%%%%%%%%%%%%%%%%%%%%%%%%%%
%    Event selections                   %
%%%%%%%%%%%%%%%%%%%%%%%%%%%%%%%%%%%%%%%%%
\begin{figure}[!h]
  %\centering
  \begin{overpic}[width=0.40\textwidth]{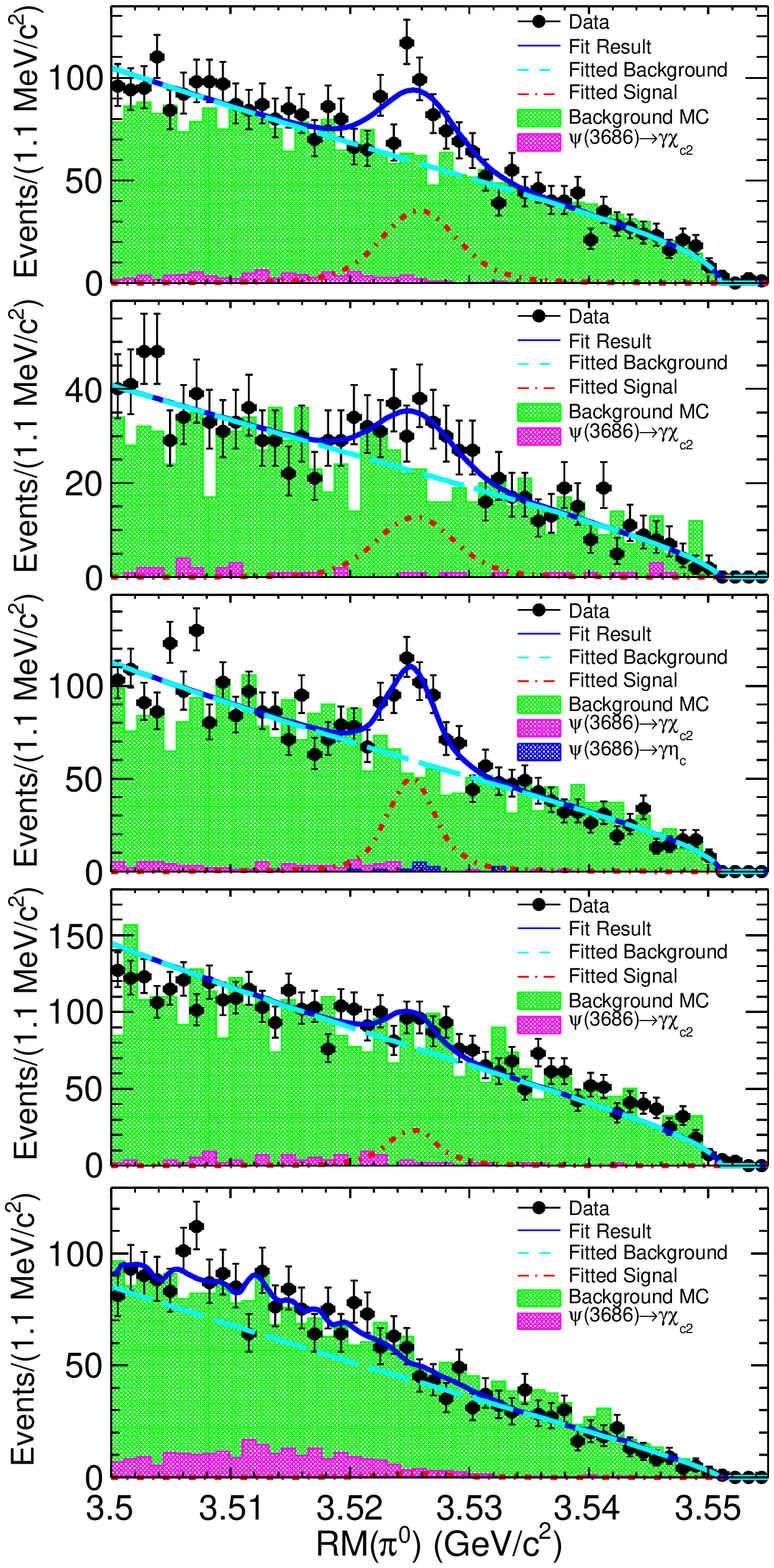}
		\put(8, 84){\bf(I)}
    \put(8, 66){\bf(II)}
    \put(8, 47){\bf(III)}
    \put(8, 28){\bf(IV)}
    \put(8, 10){\bf(V)}
  \end{overpic}
  \caption{Recoiling mass spectra of the lowest energy $\piz$, in the decay chains $\psip\to\piz\hc$ with $\hc\to
 p \pb \pip \pim$ (I), $\pip \pim \piz$ (II),  $2(\pip \pim) \piz$ (III), $3(\pip \pim) \piz$ (IV), and $\kp \km \pip \pim$
(V). In each spectrum, the dots with error bars represent data, the pink shaded histogram is the background process $\psip\to\gamma\chi_{c2}$, the blue filled histogram is the background process $\psip \to \piz \hc, \hc\to \gamma \etac$, the green filled histogram is the background from inclusive MC, the cyan dashed curve is the fitted background, the red dash-dot curve is the fitted signal, and the blue curve is the fitted result
  \ (color online).}
  \label{fig_fit}
\end{figure}

% Charged track selections
\par {\color{black}{In the following, we denote decay modes $\psip \to \piz \hc$ with $\hc \to p \pb \pip \pim,\ \pip \pim \piz,\ 2(\pip \pim) \piz,\ 3(\pip \pim) \piz$, and $\kp \km \pip \pim$ as modes I, II, III, IV, and V, respectively.
Events are selected with the expected number of charged particle candidates, and at least two photon candidates for modes I and V, and four for modes II, III, and IV.}}
% Charged track selections in MDC
Each charged track reconstructed in the MDC is required to be within 10\,cm of the interaction point along the beam direction and 1\,cm in the plane perpendicular to the beam. The polar angle $\theta$ of the tracks must be within the fiducial volume of the MDC ($|\cos\theta|<0.93$).
% PID
The TOF and $dE/dx$ information of each charged track is used to calculate the corresponding probabilities of the hypotheses that a track is a pion, kaon or proton for particle identification.
% Photon reconstruction
Electromagnetic showers are reconstructed by clustering energies deposited in the EMC, and in the nearby TOF counters. A photon candidate is such a shower with a deposited energy larger than 25\,MeV in the barrel region ($|\cos\theta|<0.8$) or 50 MeV in the end cap region ($0.86<|\cos\theta|<0.92$). The time $t$ measured in the EMC with respect to the start of the event is required to be $0 < t < 700$ ns, to suppress electronic noise and beam-associated background. The angle between the photon and the extrapolated impact point in the EMC of the nearest charged track must be larger than $10^{\circ}$ for charged pions and $20^{\circ}$ for protons, respectively, to ensure that the cluster is not from that track.

% Kinematic fits
\par Following the application of a vertex fit that constrains all the charged tracks to arise from a common interaction point, a kinematic fit is then performed to further improve resolution and suppress background. The kinematic fit applies  constraints on the four-momentum conservation between initial and final states, and imposes the nominal $\piz$ mass~\cite{pdg-2016} on $\gamma \gamma$  pairs within the interval $107 < M(\gamma\gamma) < 163$ MeV/$c^{2}$).
If there is an excess of photon candidates in the event, then all combinations are considered and the one with the smallest $\chisq$ is kept.
The $\chisq$ is required to be less than a specific value determined by maximizing $S/\sqrt{S+B}$, which is considered as a figure of merit (FOM). Here, $S$ is the number of signal events from MC simulation normalized to the preliminary result measured with the un-optimized selection criteria and $B$ is the number of background events extracted from the inclusive MC sample. The FOM is maximized in the $\hc$ signal region $|RM(\piz) - 3.525| < 8$ MeV/$\csq$, where $RM(\piz)$ is the recoiling mass of the $\piz$ meson, with the lower energy candidate chosen  in the case of multiple $\piz$s in the event.

% psi' -> gamma chic2
\par To suppress contamination from decays with different numbers of photons to the signal modes, such as the dominant background decay $\psip \to \gamma \chi_{c2}$, where the $\chi_{c2}$ decays to the same final states as the $\hc$, $\chisq_{4C.{\rm{exp}}} < \chisq_{4C.{\rm{unexp}}}$ is required for each decay mode. Here $\chisq_{4C.{\rm{exp}}}$ is obtained from the four-momentum kinematic fit that includes the expected number of photons in the signal candidate, {\color{black}{$i.e.$ two for modes I and V, and four for modes II, III, and IV}}, while $\chisq_{4C.{\rm{unexp}}}$ is obtained from a fit including an unexpected number of photons, {\color{black}{$i.e.$, one for modes I and V, and three for modes II, III, and IV}}.

%%%%%%%%%%%%%%%%%%%%%%%%%%%%%%%%%%%%%%%%%
% Background study                      %
%%%%%%%%%%%%%%%%%%%%%%%%%%%%%%%%%%%%%%%%%
% Mass windows
\par
Mass windows, optimized simultaneously with the FOM, are applied to suppress the  background contributions  from $\psip$ decays to $\piz \omega$, $\piz \eta$, $\piz \piz \jpsi$ and $\pip \pim \jpsi$, and are listed in Table~\ref{tab_vetoes}.  The residual contamination is estimated with the inclusive MC sample.

\begin{table}[h!]
  \centering\color{black}
  \caption{Mass windows imposed in background rejection.
  $M$ denotes the invariant mass $\sqrt{p^{2}}$, where $p$ is the $\pip\pim\piz$ four momentum.
  $RM$ denotes the recoiling mass $\sqrt{(p_{\psip} - p)^{2}}$, where $p_{\psip}$ is the $\psip$ four momentum, and $p$ is the $\pip\pim$, $\piz \piz$, or $\piz$ four momentum.
  $m$ denotes the nominal mass~\cite{pdg-2016} of the indicated particle.
  $\pizl$ ($\pizh$) denotes the $\piz$ candidate with lower (higher) energy.}
 % The mass window $M(\pip\pim\piz)$ is to suppress background events with $\eta \to \pip \pim \piz$ or $\omega \to \pip \pim \piz$, $RM(\pizh)$ is to suppress background events with $\psip \to \piz \omega$, while $RM(\pip\pim)$ and $RM(\pizl\pizh)$ are to suppress background events with $\psip \to \piz \piz \jpsi$ and $\psip \to \pip \pim \jpsi$, respectively.}
  \label{tab_vetoes}
  \begin{tabular}{cc}
    \hline
    \hline
    Mode &Mass windows (MeV/$\csq$)               \\
    \hline
    \multirow{3}{*}{I}    &$|RM(\pip \pim) - m(\jpsi)| > 18$        \\
         &$|M(\pip \pim \piz) - m(\eta)| > 14$     \\
         &$|M(\pip \pim \piz) - m(\omega)| > 6$   \\
    \hline
    \multirow{2}{*}{II}   &$|RM(\pizl \piz_{h}) - m(\jpsi)| > 74$   \\
         &$|RM(\piz_{h}) - m(\omega)| > 32$        \\
    \hline
    \multirow{4}{*}{III}  &$|RM(\pizl \piz_{h}) - m(\jpsi)| >20$    \\
         &$|RM(\pip \pim) - m(\jpsi)| > 22$        \\
         &$|M(\pip \pim \pizl) - m(\eta)| > 16$    \\
         &$|M(\pip \pim \pizl) - m(\omega)| > 20$  \\
    \hline
    \multirow{3}{*}{IV}   &$|RM(\pizl \piz_{h}) - m(\jpsi)| >18$    \\
         &$|RM(\pip \pim) - m(\jpsi)| > 20$       \\
         &$|M(\pip \pim \pizl) - m(\eta)| > 16$    \\
    \hline
    \multirow{3}{*}{V}    &$|RM(\pip \pim) - m(\jpsi)| > 22$        \\
         &$|M(\pip \pim \piz) - m(\eta)| > 16$     \\
         &$|M(\pip \pim \piz) - m(\omega)| > 20$   \\
    \hline
    \hline
  \end{tabular}
\end{table}

% Background study
\par Fig.~\ref{fig_fit} shows the recoiling mass distribution of $\pi^0_{l}$, the lowest energy $\piz$ candidate, obtained by applying the above selection criteria.
% data
 Clear $\hc$ signals are observed in the modes $\hc \to p \pb \pip \pim$, $\pip \pim \piz$, and $2(\pip \pim) \piz$, while no obvious signal is observed for $\hc \to 3(\pip \pim) \piz$ and $\kp \km \pip \pim$.
% Inclusive MC
For the decay mode $\hc \to 2(\pip \pim)\piz$, there are $11.0 \pm 3.3 \pm 2.5$ peaking background events from $\psip \to \piz \hc, \, \hc\to \gamma \etac$, where the first uncertainty is statistical and the second systematic, while no peaking background is found for the other decay modes, based on inclusive MC. The remaining background from $\psip \to \gamma \chi_{c2}$ is negligible for all the decay modes except $\hc \to \kp \km \pip \pim$, which will therefore be considered separately in the fit below. The background contributions from the continuum processes are studied with a 44\,pb$^{-1}$ data set taken at $\sqrt{s}=3650$ MeV, which yields no $\hc$ candidates in any of the final states analyzed.

%%%%%%%%%%%%%%%%%%%%%%%%%%%%%%%%%%%%%%%%%%%%%%%
% Fits, branching fractions, upper limits     %
%%%%%%%%%%%%%%%%%%%%%%%%%%%%%%%%%%%%%%%%%%%%%%%
\begin{table*}[hbtp]
  \centering
  \caption{Results of the analysis. Here $\epsilon$ denotes the selection efficiency, $N_{\hc}$ denotes the $\hc$ signal yield, $\br_{\psip}$ and $\br_{\hc}$ denote the branching fraction $\br(\psip \to \piz \hc)$ and $\br(\hc \to {\textrm{hadrons}})$, respectively, S.S. is the significance of the signal peak, including systematic uncertainties, and $\br_{\hc}^{\textrm{PDG}}$ denotes the branching fraction of $\hc \to {\rm hadrons}$ from the PDG~\cite{pdg-2016}. Only statistical uncertainties are presented for signal yields, while for the (product) branching fractions, the first uncertainty is statistical and the second systematic. For the decay mode $\hc \to 3(\pip \pim)\piz$  both the branching fraction and upper limit are listed.}
  \label{tab_result}
  \begin{tabular}{clcccccc}
    \hline
    \hline
% GW: introduce a little space on the left hand side of the SS column, which looked very cramped
% GW: also introduce explicit listing of each decay mode
    Mode & &$\epsilon$(\%) &$N_{hc}$ &$\br_{\psip}\times\br_{\hc}(10^{-6})$ &$\br_{\hc}(10^{-3})$ &\hspace*{0.2cm} S.S. \hspace*{0.1cm} &$\br_{\hc}^{\textrm{PDG}}(10^{-3})$ \\
    \hline
    I    &$\hc \to p \pb \pip \pim$          &20.9                     &$230\pm25$  &$2.49\pm0.27\pm0.28$  &$2.89\pm0.32\pm0.55$   &\hspace*{0.2cm}7.4$\sigma$   &-- \\
    II   &$\hc \to \pip \pim \piz$            &16.8                     &$101\pm25$  &$1.38\pm0.35\pm0.17$  &$1.60\pm0.40\pm0.32$   &\hspace*{0.2cm}4.6$\sigma$   &$<2.2$ \\
    III  &$\hc \to   2(\pip \pim) \piz$       &9.1                      &$254\pm32$  &$6.40\pm0.81\pm0.87$  &$7.44\pm0.94\pm1.52$   &\hspace*{0.2cm}9.1$\sigma$   &$22^{+8}_{-7}$ \\
    \multirow{2}{*}{IV} & \multirow{2}{*}{$\hc \to 3(\pip \pim) \piz$} & \multirow{2}{*}{4.2}     &$73\pm34$   &$4.00\pm1.87\pm0.70$  &$4.65\pm2.17\pm1.08$   &\hspace*{0.2cm}2.1$\sigma$   &\multirow{2}{*}{$<29$} \\
           &             &                         &$<136$      &$<7.5$                &$<8.7$                 & --           & \\
    V      & $\hc \to \kp \km \pip \pim$              &18.1                     &$<$40       &$<0.5$                &$<0.6$                 & --           & -- \\
    \hline
    \hline
  \end{tabular}
\end{table*}

% Fits
\par To obtain the number of signal events, an unbinned maximum likelihood fit is performed to the corresponding mass spectrum, as shown in Fig.~\ref{fig_fit}. In each fit, the signal is described with the MC simulated shape convoluted with a Gaussian function, and the background is described  with an ARGUS function~\cite{argus_func}, except for the mode $\hc \to \kp \km \pip \pim$ where an additional background component from $\psip \to \gamma \chi_{c2},\, \chi_{c2} \to \kp \km \pip \pim$ is included. Here, the MC shape includes the intrinsic $\hc$ line shape and detection resolution, while the Gaussian function accounts for the discrepancy between data and MC simulation in the mass resolution. All the parameters of the Gaussian and ARGUS functions, except the threshold  value of 3551\,MeV/$c^{2}$, are floated in the fit.

% branching fractions
\par Branching fractions are calculated based on the formula:
\begin{equation}
  \br_{\hc} = \frac{N_{\hc}}{\br(\psip \to \piz \hc) \cdot \br(\piz \to \gamma \gamma) \cdot N_{\psip} \cdot \epsilon},
  \label{eq-br}
\end{equation}
where $\br_{\hc}$ represents the branching fraction of the given signal mode,
%$\hc \to p \pb \pip \pim$, $\pip \pim \piz$, $2(\pip \pim) \piz$, $3(\pip \pim) \piz$ or $\kp \km \pip \pim$,
while $\br(\psip \to \piz \hc)$ and $\br(\piz \to \gamma \gamma)$ are the branching fractions of $\psip \to \piz \hc$ and $\piz \to \gamma \gamma$, respectively, $N_{\hc}$ and $N_{\psip}$ are the numbers of $\hc$ signal and $\psip$ events, respectively, and $\epsilon$ is the selection efficiency obtained from signal MC simulation.
% Upper limits
Since no significant signal is observed in the decays $\hc\to$ $\kp\km\pip\pim$ and $3(\pip\pim)\piz$, their upper limits are determined with a Bayesian method~\cite{bayesian01}. With the fit function described before, we scan the number of signal yield to obtain the likelihood distribution, and smear it with the systematic uncertainties. The upper limits of the number of signal yield $N^{up}_{\hc}$ at the 90\% confidence level are obtained via $\int^{N^{up}_{\hc}}_{0}F(x)dx/\int^{\infty}_{0}F(x)dx=0.90$, where $F(x)$ is the probability density function of the likelihood distribution. All the numerical results, including selection efficiencies, signal yields, branching fractions or upper limits and significances, are listed in Table~\ref{tab_result}.

%%%%%%%%%%%%%%%%%%%%%%%%%%%%%%%%%%%%%%%%%%%%%%%
% Systematic uncertainties                    %
%%%%%%%%%%%%%%%%%%%%%%%%%%%%%%%%%%%%%%%%%%%%%%%
\begin{table}[h!]
  \centering
  \caption{Relative uncertainties (in \%) on the branching fractions.}
  \begin{tabular}{lp{0.8cm}<{\centering}p{0.8cm}<{\centering}p{0.8cm}<{\centering}p{0.8cm}<{\centering}p{0.8cm}<{\centering}}
  %\begin{tabular}{lccccc}
    \hline
    \hline
    Source &I &II &III &IV &V \\
    \hline
    Tracking              &5.0 &2.0 &4.0 &6.0 &4.0\\
    Photon                &2.0 &4.0 &4.0 &4.0 &2.0 \\
    $\piz$ reconstruction &1.0 &2.0 &2.0 &2.0 &1.0 \\
    PID                   &4.9 &2.0 &4.0 &6.0 &4.0 \\
    Kinematic fit         &1.8 &2.2 &3.7 &4.2 &1.5 \\
    Number of $\psip$     &0.7 &0.7 &0.7 &0.7 &0.7 \\
    Fitting range         &2.6 &3.5 &4.9 &--  &-- \\
    Signal shape          &1.3 &8.1 &2.5 &--  &-- \\
    Background shape      &2.1 &3.5 &2.9 &--  &-- \\
    Resolution            &4.2 &5.1 &3.3 &--  &-- \\
    $\etac$               &--  &--  &1.5 &--  &-- \\
    Physics model         &6.3 &2.6 &8.2 &14.1 &7.3 \\
    Sum                   &11.3 &12.5 &13.6 &17.6 &9.6 \\
    \hline
    \hline
  \end{tabular}
  \label{tab_sys_err}
\end{table}

% Systematic uncertainties
\par The sources of systematic uncertainties for the product branching fractions include tracking, photon and $\piz$ reconstruction, PID, the kinematic fit, the number of $\psip$ events, fitting procedure, $\etac$ peaking background, mass windows and the physics model describing the $\hc$ production and decay dynamics. All the systematic uncertainties are summarized in Table~\ref{tab_sys_err}, and the overall systematic uncertainties are obtained by summing all individual components in quadrature. In addition, we add a relative systematic uncertainty of 15.2\% assocated with the branching fraction of $\psip \to \piz \hc$ in calculating the branching fraction of the $\hc$ hadronic decays.

% Tracking and photon
\par The uncertainties on the tracking efficiency are estimated with the control samples $\psip\to\pip\pim\jpsi$, $\jpsi\to K_{S}^{0}K^{\pm}\pi^{\mp}$ and $\psip\to\p\pb\pip\pim$, and are determined to be 1.0\%~\cite{tracking_pion}, 1.0\%~\cite{tracking_kaon}, 1.3\%, and 1.7\% for each charged pion, kaon, proton, and anti-proton, respectively. The uncertainties on the photon and $\piz$ reconstruction efficiency are studied using the control sample $\jpsi \to \pip \pim \piz$, and are determined to be 1.0\% per photon~\cite{photon_rec} and 1\% per $\piz$~\cite{photon_rec}.
% PID
The PID uncertainties are determined to be 1.0\% per pion~\cite{pid_pion}, 1.0\% per kaon~\cite{tracking_kaon}, 1.3\% per proton and 1.6\% per antiproton, based on the same samples used to estimate tracking uncertainties.
% Kinematic fit
The uncertainty associated with the kinematic fit is estimated by comparing the efficiencies with and without the helix parameter correction~\cite{kine_fit01}.
% psi' number
The uncertainty on the number of $\psip$ events is 0.7\%, according to the study in Ref.~\cite{psip-numerr}.

% Fitting procedure
\par The fitting range, signal and background descriptions, and the difference in resolution between data and simulation are considered as sources of systematic uncertainty related to the fitting procedure. These uncertainties are assigned by varying the boundaries of the fitting ranges by $\pm$10 MeV/$c^{2}$, changing the signal description from the shape determined from the simulation to a Breit-Wigner function, and replacing the ARGUS function describing the background with a second-order Chebychev polynomial.
The difference between the results obtained by fixing and releasing the resolution in the fit is taken as the uncertainty on the knowledge of this quantity, where in the former case a correction of 1\,MeV/$c^2$ is first applied to the value from the simulation, as determined from a  control sample $\psip\to\gamma\chi_{c1}\to\gamma\p\pb\pip\pim$.
For $\hc\to3(\pip\pim)\piz$ and $\kp\km\pip\pim$, the largest upper limits are taken with different combinations of fitting models and ranges.
% =============
The uncertainty due to $\etac$ peaking background  is assigned from the statistical uncertainty on the fit result for this component, and the corresponding uncertainty on the branching fractions.
% Mass windows
% GW: comment out below line,  as authors say that this uncertainty will be removed.  Table III will have to be adjusted accordingly when they do so.
%Another uncertainty comes from the assigned ranges of the mass windows imposed for the background veto, which is determined by varying these ranges.

\par A systematic uncertainty due to  the physics model arises from the limited knowledge of the intermediate states in $\hc$ decays.
% Intermediate states
Searches have been performed for intermediate states contributing to modes  I to III, which are detailed in the Supplemental Material~\cite{supple}.  Possible contributions are found for several such states, which including a $\rho^{0}$ peak in each projection of the $\pip\pim$ invariant mass.   The effect of these states on the selection efficiency  is evaluated by generating alternative simulation samples with different properties and comparing with the default production.

%%%%%%%%%%%%%%%%%%%%%%%%%%%%%%%%%%%%%%%%%%%%%%%
% Summary                                     %
%%%%%%%%%%%%%%%%%%%%%%%%%%%%%%%%%%%%%%%%%%%%%%%

\begin{table}[hbtp]
  \caption{The ratios of the hadronic decay widths of $\hc$ to $\etac$ ($\Gamma_{\hc}^{\rm had} / \Gamma_{\etac}^{\rm had}$) and $\hc$ to $\jpsi$ ($\Gamma_{\hc}^{\rm had} / \Gamma_{\jpsi}^{\rm had}$). The theoretical predictions of the total hadronic decay ratios are based on pQCD and NRQCD~\cite{theory-1}, which are expected to be correct also for exclusive decay modes. The experimental measurements of the ratios of the partial decay widths for $p \pb \pip \pim$, $\kp\km\pip\pim$, and $n(\pip \pim)\piz (n = 0, 1, 2)$ modes are calculated based on the measured branching fractions in this analysis and the PDG~\cite{pdg-2016}.}
  \label{ratios}
  \begin{tabular}{c|cc}%p{0.8cm}<{\centering}p{0.8cm}<{\centering}p{0.8cm}<{\centering}p{0.8cm}<{\centering}p{0.8cm}<{\centering}}
    \hline
    \hline
    &Model/Mode &Ratio \\
    \hline
    \multirow{4}{*}{$\Gamma_{\hc }^{\rm had}/\Gamma_{\etac }^{\rm had}$} &pQCD  &$0.010\pm0.001$ \\
    &NRQCD               &$0.083\pm0.018$ \\
    \cline{2-3}
    &$p \pbar \pip \pim$ &$0.012 \pm 0.008$ \\
    &$\kp \km \pip \pim$ &$<0.083$ \\
    \hline
    \multirow{7}{*}{$\Gamma_{\hc }^{\rm had}/\Gamma_{\jpsi}^{\rm had}$} &pQCD &$0.68\pm0.07$ \\
    &NRQCD                &$8.03\pm1.31$ \\
    \cline{2-3}
    &$p \pbar \pip \pim$  &$3.63 \pm 2.25$ \\
    &$\pip \pim \piz$     &$0.57 \pm 0.38$ \\
    &$2(\pip \pim) \piz$  &$1.43 \pm 0.90$ \\
    &$3(\pip \pim) \piz$ &$<2.26$ \\
    &$\kp \km \pip \pim$ &$<0.68$ \\
    \hline
    \hline
  \end{tabular}
\end{table}

\par In summary, three $h_c$ hadronic decays, $\hc\to\p\pb\pip\pim$, $\hc\to\pip\pim\piz$, and $\hc\to2(\pip\pim)\piz$, are observed for the first time, and two channels, $\hc\to\kp\km\pip\pim$ and $\hc\to3(\pip\pim)\piz$, are searched for.
The measured branching fractions or upper limits, as well as the significance of the signal peaks, are listed in Table~\ref{tab_result}.
The measured branching fraction of $\hc\to 2(\pip \pim) \piz$ is more precise than the CLEO-c result~\cite{Adams:2009aa} and lower in value, although consistent within uncertainties.
The sum of the branching fractions of the three observed channels is approximately $1.2\%$, which is still smaller than the $h_c$ radiative transition to the $\eta_c$, and does not yet allow a conclusion on whether the  total hadronic decay width of the $h_c$ is of the same order as its radiative transition.
%So, measurements of more $\hc$ hadronic decay modes should be helpful to clarify this problem.
Table~\ref{ratios} shows the comparisons of the measured ratios of the hadronic decay widths $\Gamma_{\hc}^{\rm had} / \Gamma_{\etac}^{\rm had}$ and $\Gamma_{\hc}^{\rm had} / \Gamma_{\jpsi}^{\rm had}$ and the theoretical predictions.
The experimental results tend to favour the lower predictions, which come from pQCD.
However, in Ref.~\cite{theory-1}, the theoretical prediction of $\br (\hc \to \gamma \etac) = (41 \pm 3)\%$ based on NRQCD is favored by the experimental measurement $(51\pm 6) \%$~\cite{pdg-2016}, compared with the prediction of $(88 \pm 2)\%$ from pQCD.
We note that the experimental measurements are still limited by low statistics and the predictions of the theoretical models can be modified through considerations such as normalization scale or relativistic corrections~\cite{Zhang:2014qqa,Li:2012rn}. Future experimental measurements of higher precision, and improved theoretical calculations will help to resolve this inconsistency.

%%%%%%%%%%%%%%%%%%%%%%%%%%%%%%%%%%%%%%%%%%%%%%%
% acknowledge, references                     %
%%%%%%%%%%%%%%%%%%%%%%%%%%%%%%%%%%%%%%%%%%%%%%%
\par The BESIII collaboration thanks the staff of BEPCII and the IHEP computing center for their strong support. This work is supported in part by National Key Basic Research Program of China under Contract No. 2015CB856700; National Natural Science Foundation of China (NSFC) under Contracts Nos. 11335008, 11425524, 11625523, 11635010, 11735014, 11565006; the Chinese Academy of Sciences (CAS) Large-Scale Scientific Facility Program; the CAS Center for Excellence in Particle Physics (CCEPP); Joint Large-Scale Scientific Facility Funds of the NSFC and CAS under Contracts Nos. U1532257, U1532258, U1732263; CAS Key Research Program of Frontier Sciences under Contracts Nos. QYZDJ-SSW-SLH003, QYZDJ-SSW-SLH040; 100 Talents Program of CAS; INPAC and Shanghai Key Laboratory for Particle Physics and Cosmology; German Research Foundation DFG under Contracts Nos. Collaborative Research Center CRC 1044, FOR 2359; Istituto Nazionale di Fisica Nucleare, Italy; Koninklijke Nederlandse Akademie van Wetenschappen (KNAW) under Contract No. 530-4CDP03; Ministry of Development of Turkey under Contract No. DPT2006K-120470; National Science and Technology fund; The Swedish Research Council; U. S. Department of Energy under Contracts Nos. DE-FG02-05ER41374, DE-SC-0010118, DE-SC-0010504, DE-SC-0012069; University of Groningen (RuG) and the Helmholtzzentrum fuer Schwerionenforschung GmbH (GSI), Darmstadt.

\end{document}

%% file: authors.tex
\author{
%\begin{small}
\begin{center}
M.~Ablikim$^{1}$, M.~N.~Achasov$^{10,d}$, S. ~Ahmed$^{15}$, M.~Albrecht$^{4}$, M.~Alekseev$^{55A,55C}$, A.~Amoroso$^{55A,55C}$, F.~F.~An$^{1}$, Q.~An$^{52,42}$, J.~Z.~Bai$^{1}$, Y.~Bai$^{41}$
% ...
\\
\vspace{0.2cm}
(BESIII Collaboration)\\
\vspace{0.2cm} {\it
$^{1}$ Institute of High Energy Physics, Beijing 100049, People's Republic of China\\
% ...
$^{4}$ Bochum Ruhr-University, D-44780 Bochum, Germany\\
% ...
$^{10}$ G.I. Budker Institute of Nuclear Physics SB RAS (BINP), Novosibirsk 630090, Russia\\
% ...
$^{15}$ Helmholtz Institute Mainz, Johann-Joachim-Becher-Weg 45, D-55099 Mainz, Germany\\
% ...
$^{41}$ Southeast University, Nanjing 211100, People's Republic of China\\
$^{42}$ State Key Laboratory of Particle Detection and Electronics, Beijing 100049, Hefei 230026, People's Republic of China\\
% ...
$^{52}$ University of Science and Technology of China, Hefei 230026, People's Republic of China\\
% ...
$^{55}$ (A)University of Turin, I-10125, Turin, Italy; (B)University of Eastern Piedmont, I-15121, Alessandria, Italy; (C)INFN, I-10125, Turin, Italy\\
% ...
$^{d}$ Also at the Novosibirsk State University, Novosibirsk, 630090, Russia\\
% ...
}\end{center}
\vspace{0.4cm}
%\end{small}
(Dated: {\today})
}